\documentclass[aps,prd,showpacs,preprint]{revtex4}
\usepackage{graphicx}
\begin{document}
\title{Exotic Quark Production in ep Collisions}
\author{A. T. Alan, A. Senol and N. Karagoz}
\address{Department of Physics, Abant Izzet Baysal University,
14280, Bolu, Turkey}

\pacs{12.60.-i, 13.60.-r}
\begin{abstract}

We investigate the single production and decay of charge -1/3
,weak isosinglet vectorlike exotic $D$ quarks in string inspired
$E_6$ theories at future ep colliders; THERA with $\sqrt s$=1 TeV,
$L=40$ $pb^{-1}$ and CERN Large Hadron Electron Collider (LHeC)
with $\sqrt s$=1.4 TeV, $L=10^4$ $pb^{-1}$. We found that an
analysis of the decay modes of $D$ should probe the mass ranges of
100-450 GeV and 100-1200 GeV at the center of mass energies, 1 and
1.4 TeV , respectively.
\end{abstract}
\maketitle
\section{Introduction}

String inspired $E_6$ theories predict existence of exotic
particles. In $E_6$, each generation of fermions is assigned to
the 27-dimensional representation
\cite{Rizzo:1985kn,Hewett:1988xc}. The presence of additional
fermions causes Flavor Changing Neutral Current (FCNC)
interactions and possible deviations from weak universality in
Charged Current (CC) \cite{Barger:1985nq}. In this study we
consider exotic down quark ($D$), a charge -1/3, quark which is a
weak isosinglet particle.

Production of exotic quarks have been studied at HERA
\cite{Hewett:1987rx,Almeida:1994pw} and LEP \cite{Rizzo:1989dt}
energies as CC and FCNC reactions. We analyze the possible
production of these quarks and some of their indirect signatures
including the contributions of boson-gluon fusions
\cite{Schuler:1987wj} at two future high energy $ep$ collider
options; THERA with $\sqrt s$=1 TeV and $L=40$ $pb^{-1}$
\cite{Abramowicz:2001qt} and CERN Large Hadron Electron Collider
(LHeC) with $\sqrt s$=1.4 TeV and $L=10^4$ $pb^{-1}$ at which 7
TeV LHC protons collide with 70 GeV ring electron or positron beam
\cite{Dainton:2006wd}. These colliders complement the hadron
collider programmes and provide new discovery potential to them. A
relatively high integrated luminosity of 10 $fb^{-1}$ at the LHeC
makes an essential facility to resolve possible puzzles of the LHC
data.
\section{Production and decays of $D$}
\begin{figure}[t]
\centering
\includegraphics[width=5.5cm]{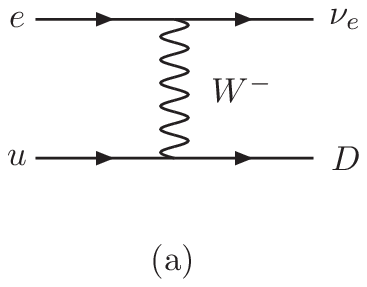}\includegraphics[width=5.5cm]{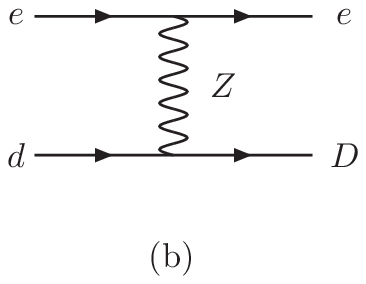}\\
\includegraphics[width=5.5cm]{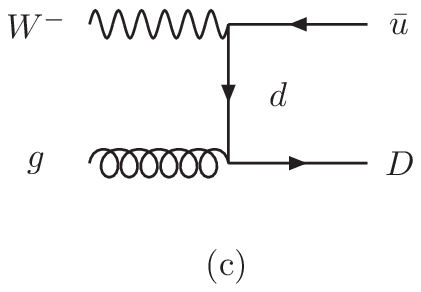}\includegraphics[width=5.5cm]{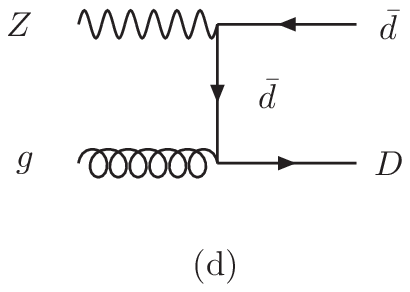}\\
\includegraphics[width=5.5cm]{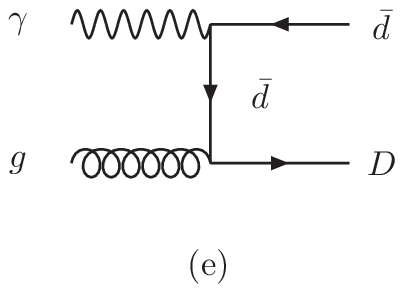}
\caption{The Feynman diagram at parton level for single exotic
down quark production in ep collisions via (a) charged current
interaction (b) neutral current interaction (c) $W$-gluon fusion
(d) $Z$-gluon fusion (e) $\gamma$-gluon fusion }\label{fig1}
\end{figure}
The single production of exotic down quarks occur via the
following t-channel subprocesses in $ep$ collisions as shown in
Fig.~\ref{fig1};\\
i) Charged Current reaction; $eu\rightarrow\nu_eD$\\
ii) FCNC reaction; $ed\rightarrow eD$\\
iii) Boson-Gluon Fusions; $Wg\rightarrow\bar uD$,
$Zg\rightarrow\bar d D$ and $\gamma g\rightarrow \bar d D$.\\
The CC and FCNC interactions for the exotic $D$ quarks mixed with
standard fermions and standard bosons $W$, $Z$ are given by
\begin{eqnarray}
\mathcal{L}_{CC}&=&\frac{g}{2\sqrt 2}\bar
u\gamma_{\mu}(1-\gamma_5)(d\cos\theta-D\sin\theta)W^{\mu}+h.c.\\
\mathcal{L}_{NC}&=&\frac{g}{4c_W}\sin\theta\cos\theta\bar
d\gamma_{\mu}(1-\gamma_5)DZ^{\mu}
\end{eqnarray}
where $\theta$ are the mixing angels between the ordinary quarks
and the exotic down quarks, $g$ denotes the gauge coupling
relative to SU(2) symmetries.

As we consider the strong interactions, $D$ quarks couple to
gluons in exactly the same way as ordinary quarks providing the
($W$, $Z$, $\gamma$)-gluon fusions. Corresponding differential
cross sections are written as follows:
\begin{eqnarray}
\frac{d\hat\sigma}{d\hat
t}(eu\rightarrow\nu_eD)&=&\frac{\pi\alpha^2\sin^2\theta}{4\sin^4\theta_W
\hat s[(\hat t-M_W^2)^2+\Gamma_W^2M_W^2]}[(\hat s-m^2)]
\nonumber\\
\frac{d\hat\sigma}{d\hat t}(ed\rightarrow
eD)&=&\frac{\pi\alpha^2\sin^22\theta[\hat t(\hat t+2\hat
s-m^2)(a_e-v_e)^2+2\hat s(\hat s-m^2)(a_e^2+v_e^2)]}{4\hat
s^2\sin^42\theta_W[(\hat t-M_Z^2)^2+\Gamma_Z^2M_Z^2]}
\nonumber\\
\frac{d\hat\sigma}{d\hat t}(Wg\rightarrow\bar
uD)&=&\frac{\pi\alpha_s\alpha\cos^2\theta}{32\hat s^2\hat
t^2M_W^2}[2M_W^4(m^2+\hat t)-2M_W^2\hat t(\hat s+\hat t)-(m^2-\hat
s)\hat t^2]
\\
\frac{d\hat\sigma}{d\hat t}(Zg\rightarrow\bar d
D)&=&\frac{\pi\alpha_s\alpha(a_d^2+v_d^2)}{8\sin^22\theta_W\hat
s^2\hat t^2M_Z^2}[2M_Z^4(m^2+\hat t)-2M_Z^2\hat t(\hat s+\hat
t)-(m^2-\hat s)\hat t^2]
\nonumber\\
\frac{d\hat\sigma}{d\hat t}(\gamma g\rightarrow \bar d
D)&=&\frac{\pi\alpha_s\alpha}{24\hat s^2\hat t}[-\hat s-\hat
t]\nonumber
\end{eqnarray}
here $M_W$, $M_Z$ and $\Gamma_W$, $\Gamma_Z$ are masses and decay
widths of a $W$ and $Z$ bosons. $a_e(a_d)$ and $v_e(v_d)$ stand
for axial and vector coupling constants of electron (down quark)
and $m$ refers to $D$ masses.

\begin{figure}
\centering
  \includegraphics[width=12cm]{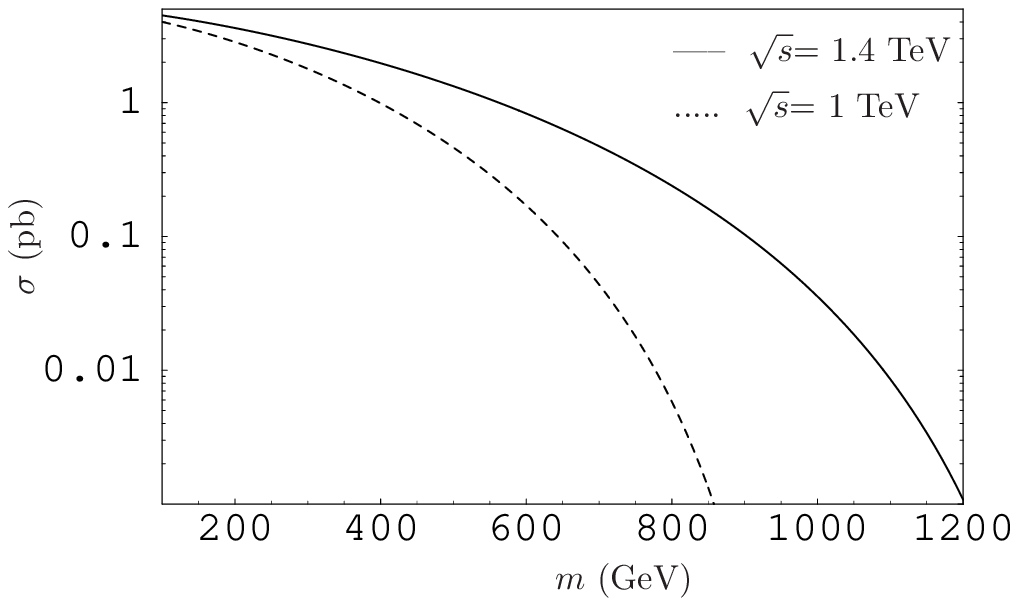}
\caption{The total production cross sections for the subprocesses
$eu\rightarrow \nu_eD$ as a functions of $m$.}\label{fig2}
\end{figure}

\begin{figure}
\centering
  \includegraphics[width=12cm]{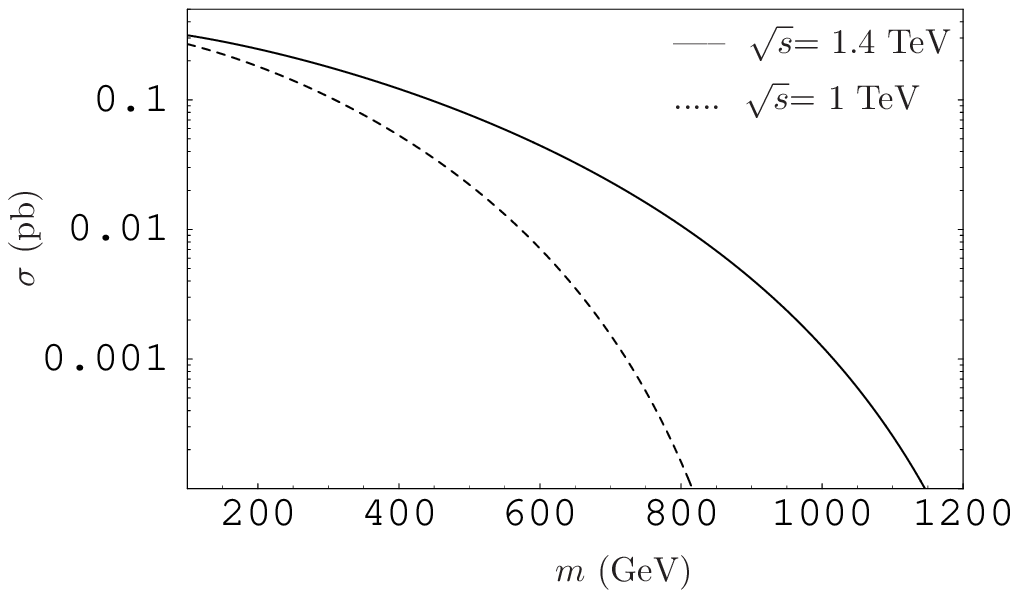}
\caption{The total production cross sections for the subprocesses
$ed\rightarrow eD$ as a functions of $m$.}\label{fig3}
\end{figure}

\begin{figure}
\centering
  \includegraphics[width=12cm]{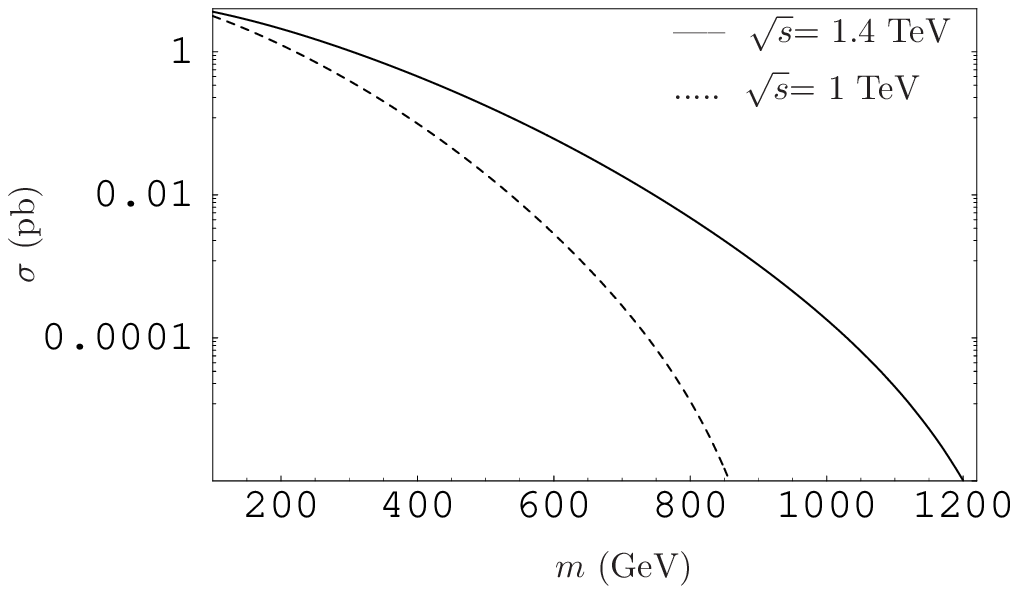}
\caption{The total production cross sections for the subprocesses
$Wg\rightarrow \bar uD$ as a functions of $m$.}\label{fig4}
\end{figure}

\begin{figure}
\centering
  \includegraphics[width=12cm]{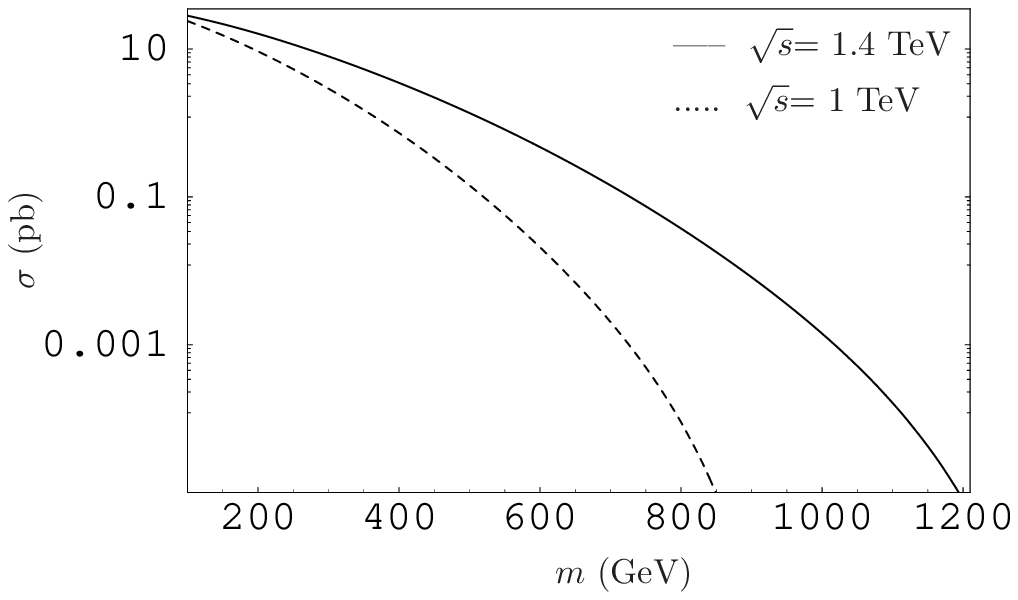}
\caption{The total production cross sections for the subprocesses
$Zg\rightarrow \bar dD$ as a function of $m$.}\label{fig5}
\end{figure}

\begin{figure}
\centering
  \includegraphics[width=12cm]{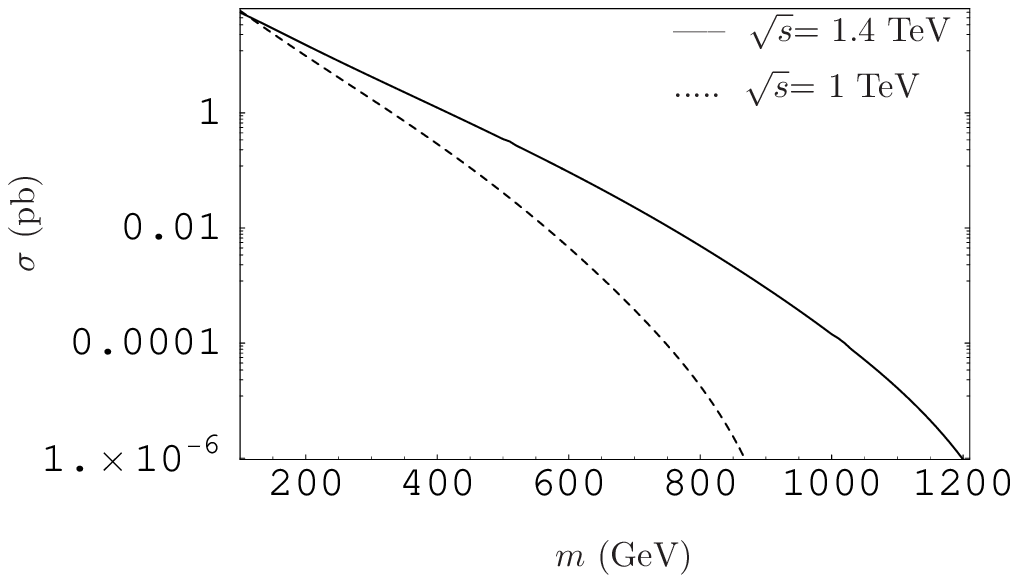}
\caption{The total production cross sections for the subprocesses
$\gamma g\rightarrow \bar d D$ as a functions of $m$.}\label{fig6}
\end{figure}

The total production cross section is obtained by folding the
partonic cross section over the parton distributions in the
proton. In numerical calculations of the total cross sections we
have used the MRST parametrization \cite{Martin:1998sq} for the
partons and the Weizs\"{a}cker-Williams distribution
\cite{vonWeizsacker:1934sx,Williams:1934ad} for photons in
electrons. For illustrations we have assumed an upper bound of
$\sin^2 \theta\lesssim 0.05$ in numerical calculations which is
appropriate as being at the order of CKM angle. The calculated
total cross sections corresponding to five subprocess have been
displayed in Figures~\ref{fig2}-\ref{fig6} as functions of $D$
masses $m$ for two center of mass energies. In these figures solid
(dashed) lines are for $\sqrt s$=1.4 (1) TeV. In Tables~\ref{tab1}
and ~\ref{tab2}, we present the total production cross sections of
the five reactions for various masses at THERA and LHeC,
respectively. As can be seen, for 100-300 GeV $D$ quarks $Z$-gluon
and $\gamma$-gluon fusions are dominant reactions but for higher
mass range contributions of these fusions decrease very fast.

Since these exotic quarks must be at a scale well above 100 GeV
\cite{Eidelman:2004wy}, the main decays of them would be
$D\rightarrow dZ$ and $D\rightarrow uW$ and partial decay widths
can be written as
\begin{eqnarray}\label{eq4}
\Gamma(D\rightarrow qV)=\alpha C_V\frac{m}{Y^2}(1-3Y^4+2Y^6)
\end{eqnarray}
where $q$ is up or down quark, $V$ denotes $W^-$ and $Z$ bosons,
$\alpha$ is the fine structure constant, $Y\equiv M_V/m$,
~$C_W=\sin^2\theta/8x_W$,~
$C_Z=(\sin\theta\cos\theta)^2/16x_W(1-x_W)$ and $x_W\equiv
\sin^2\theta_W$. Eq.~(\ref{eq4}) gives rise to branching ratios of
BR($D\rightarrow dZ$)=30 \% and BR($D\rightarrow uW$)= 70 \% which
do not change significantly depending on $D$ masses, as seen from
Table~\ref{tab3}. Therefore, $D\rightarrow ul^-\bar\nu_l$ is taken
as the relevant background process since the $D\rightarrow uW$ is
dominant one. In Tables~\ref{tab4} and~\ref{tab5} we present the
cross sections resulting the  background processes ( considering
only the first generation of leptons) for $\sqrt s$=1 and 1.4 TeV,
respectively. Results reported in these tables were obtained by
using the high energy package CompHEP \cite{Pukhov:1999gg} along
with CTEQ6L \cite{Lai:1994bb} which has been used in the
background calculations and an optimal cut of $P_T
> 10$ GeV has taken for electron, jet and missing momenta.

\begin{figure}
\centering
 \includegraphics[width=7cm]{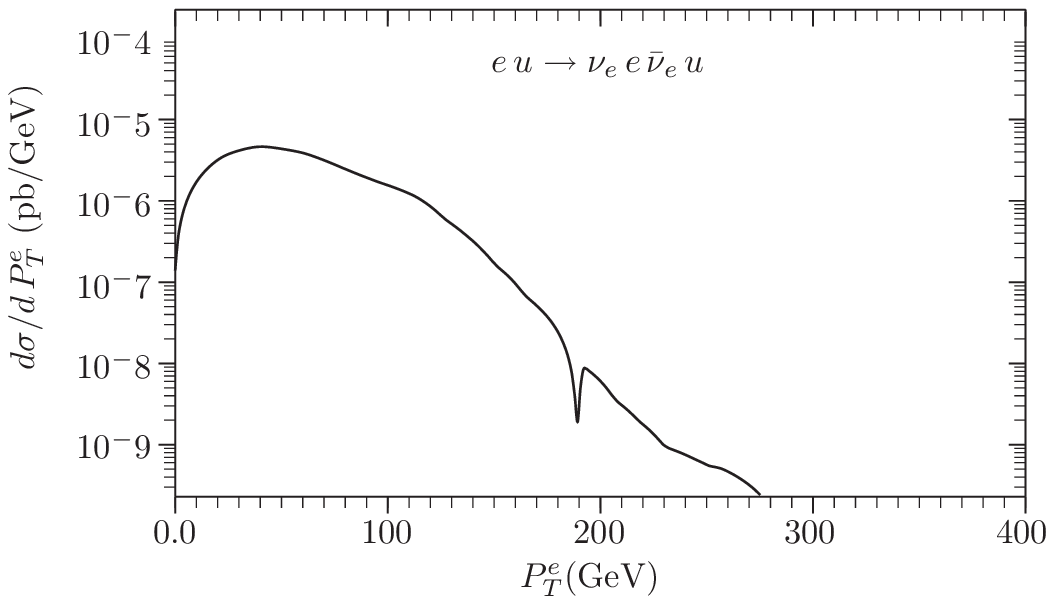}\includegraphics[width=7cm]{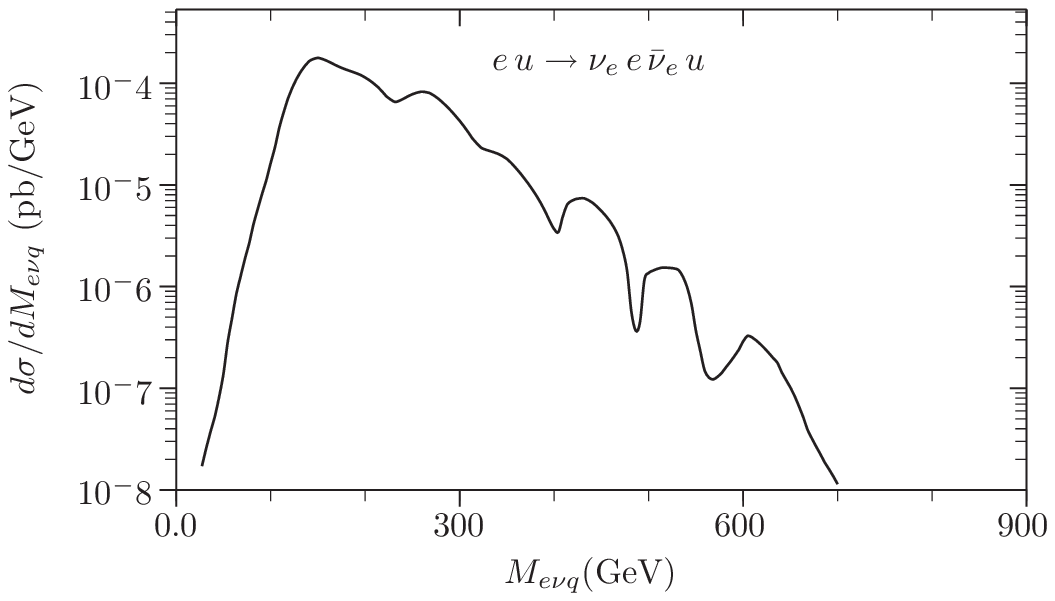}
  \includegraphics[width=7cm]{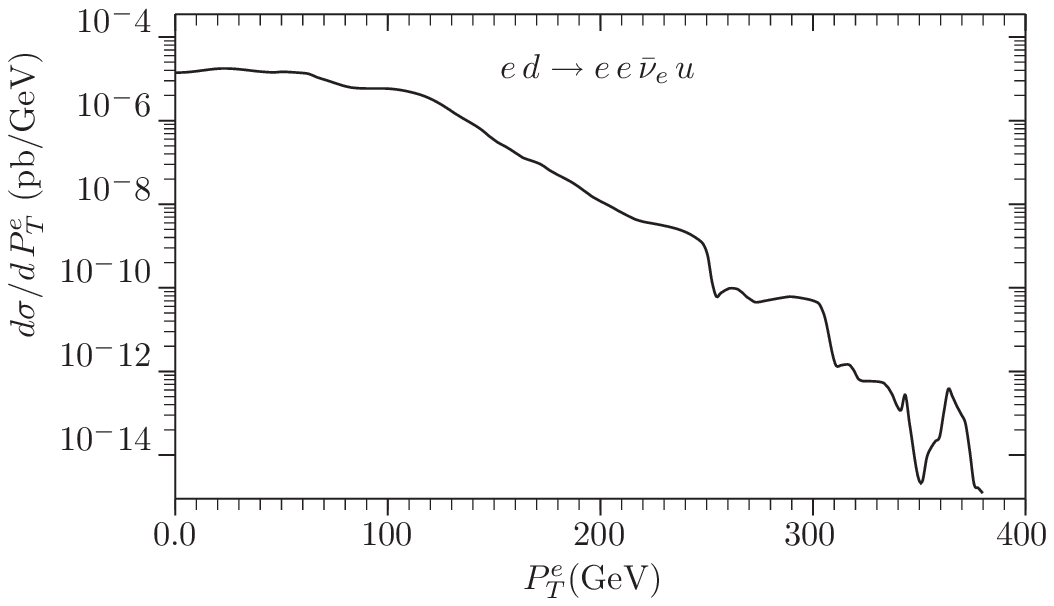}\includegraphics[width=7cm]{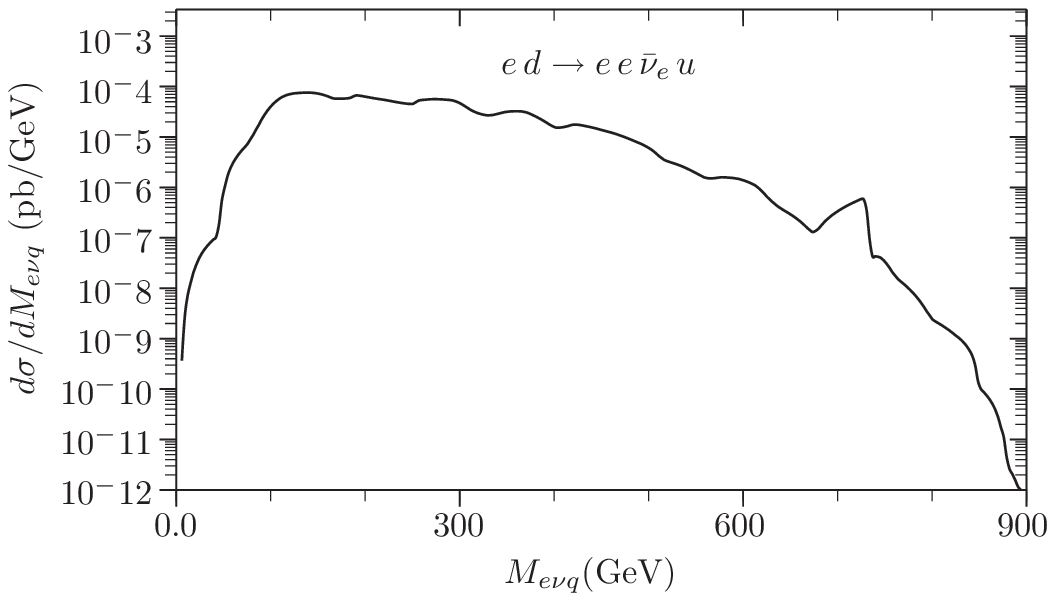}
    \includegraphics[width=7cm]{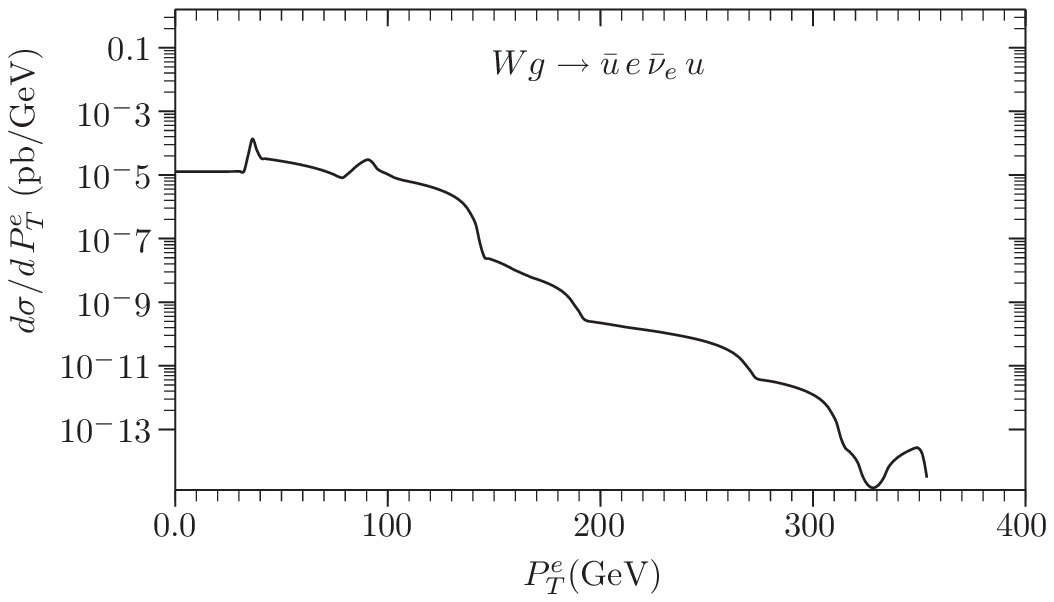}\includegraphics[width=7cm]{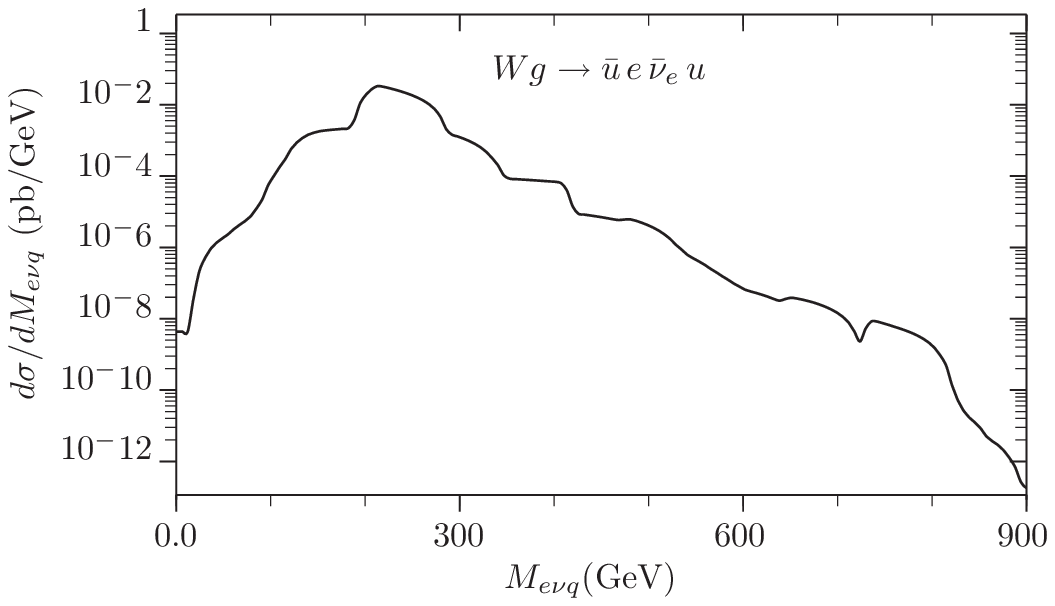}
    \includegraphics[width=7cm]{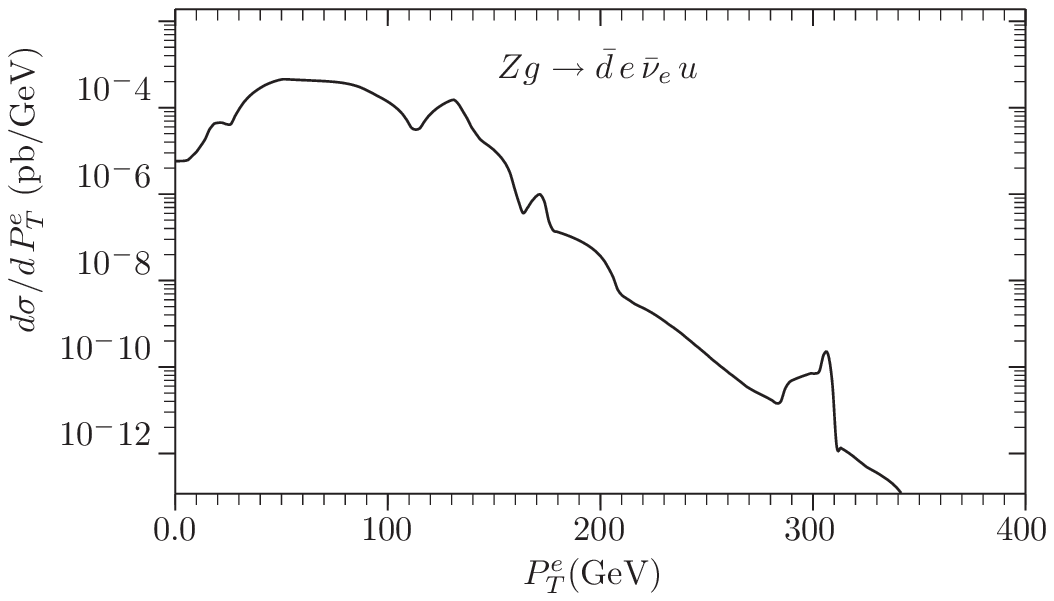}\includegraphics[width=7cm]{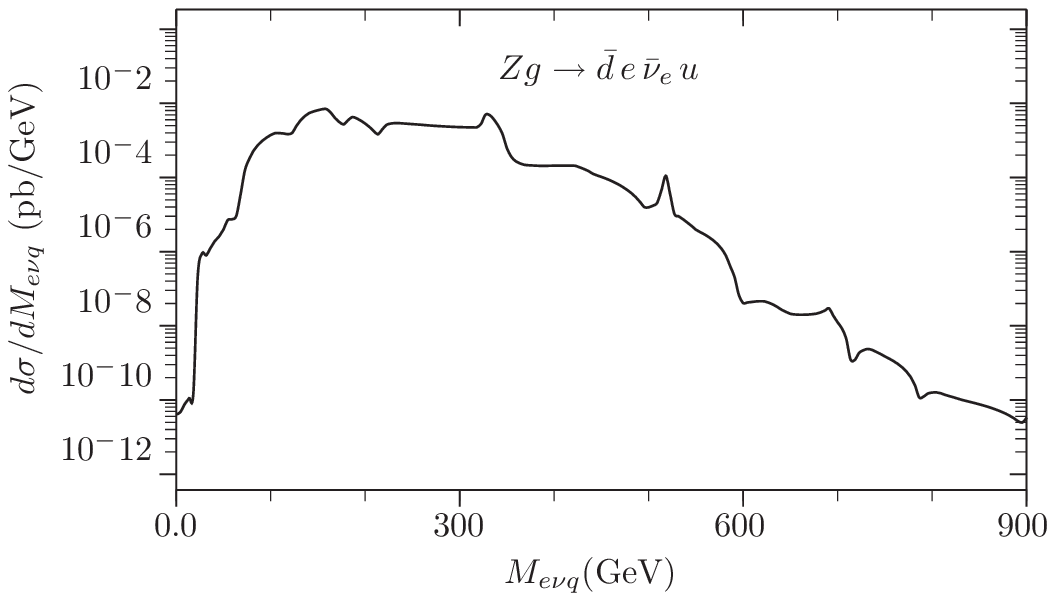}
    \includegraphics[width=7cm]{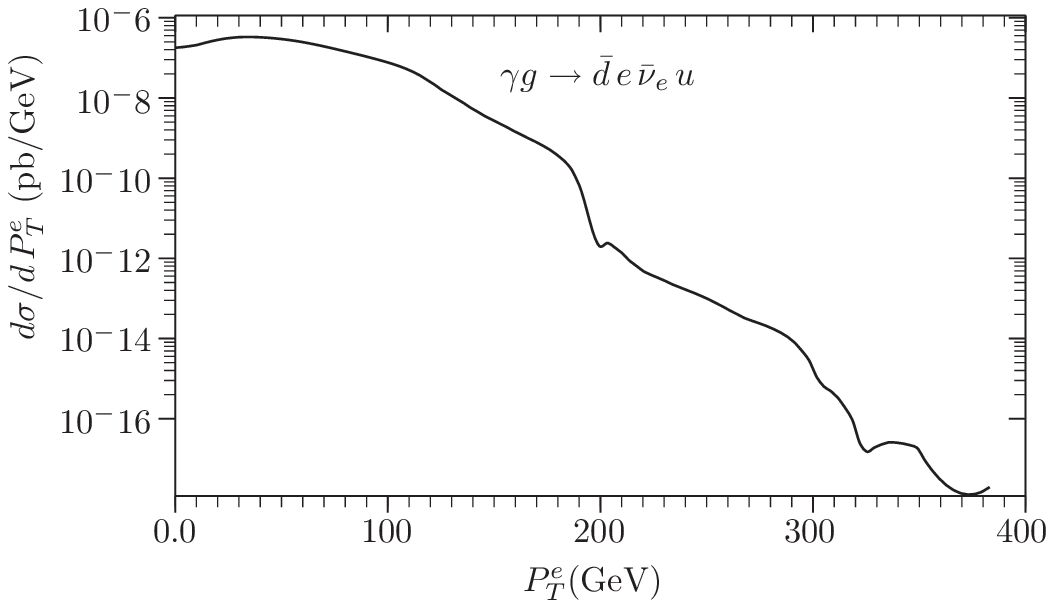}\includegraphics[width=7cm]{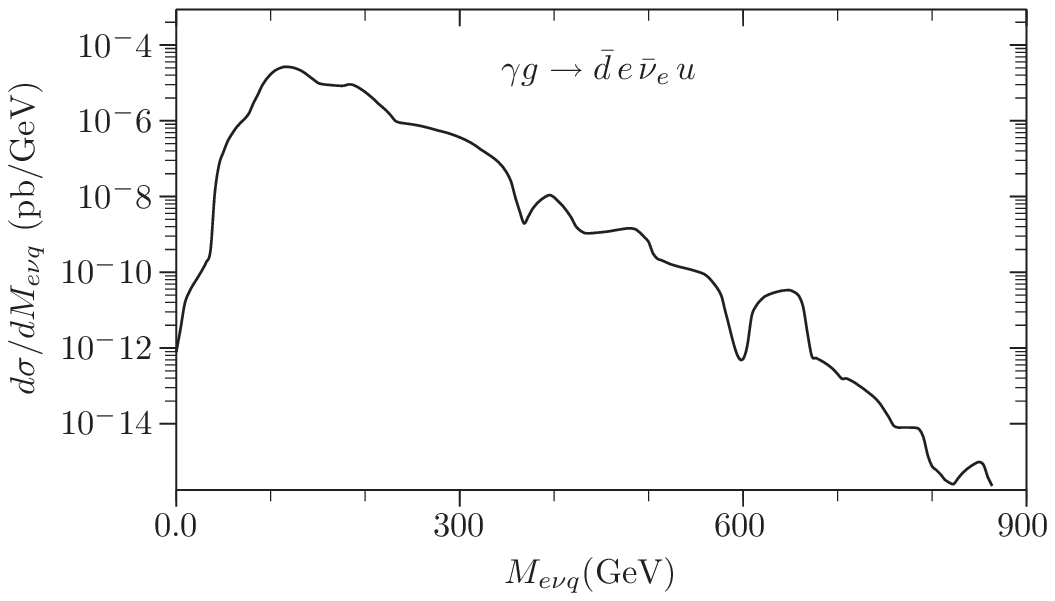}
\caption{Transverse momentum $P_T^e$ and invariant mass
distributions of the backgrounds at THERA($\sqrt{S}$=1
TeV).}\label{fig7}
\end{figure}

\begin{figure}
\centering
 \includegraphics[width=7cm]{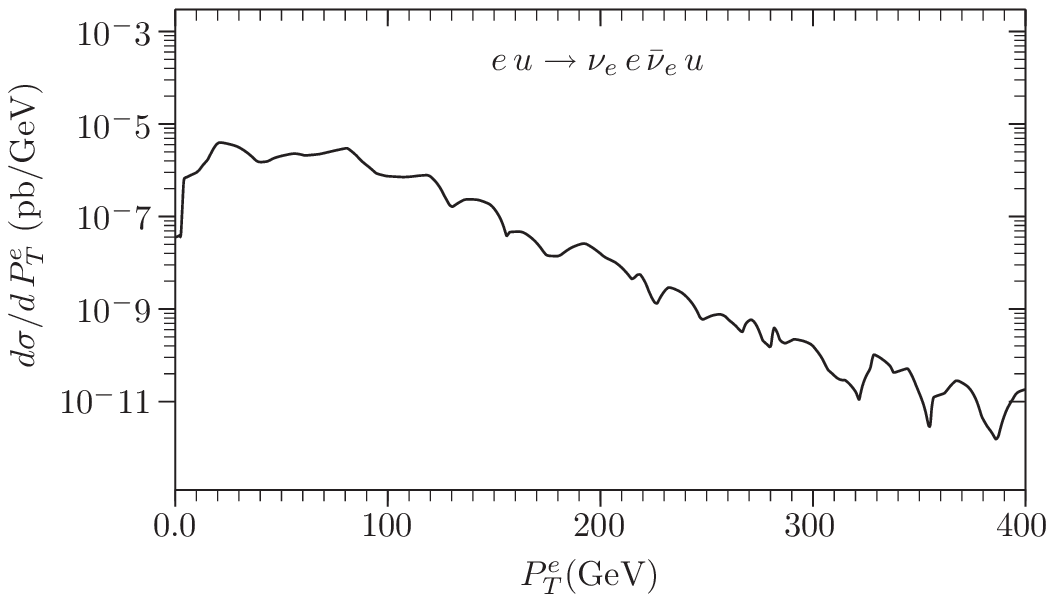}\includegraphics[width=7cm]{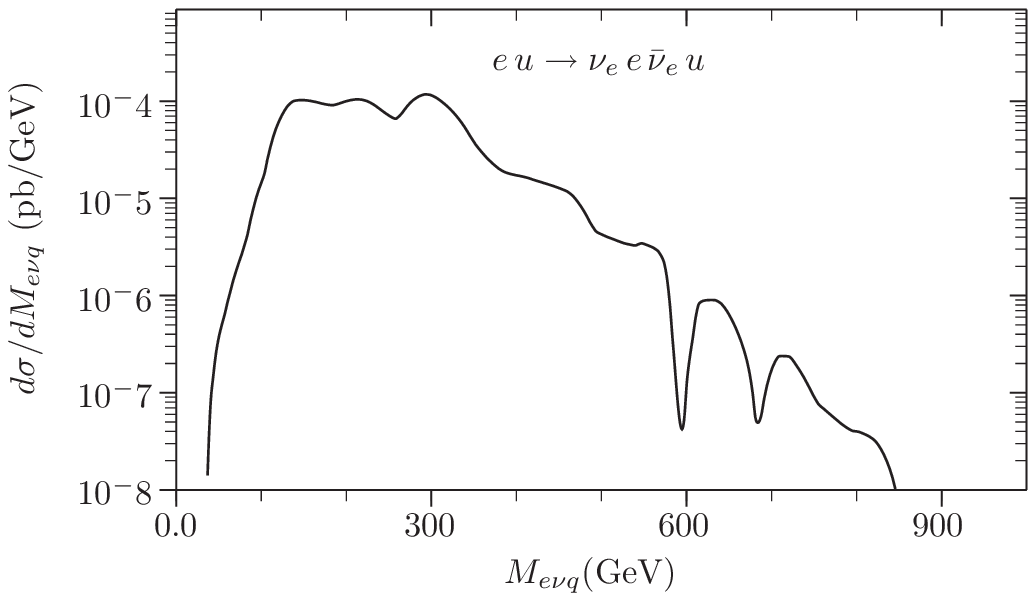}
  \includegraphics[width=7cm]{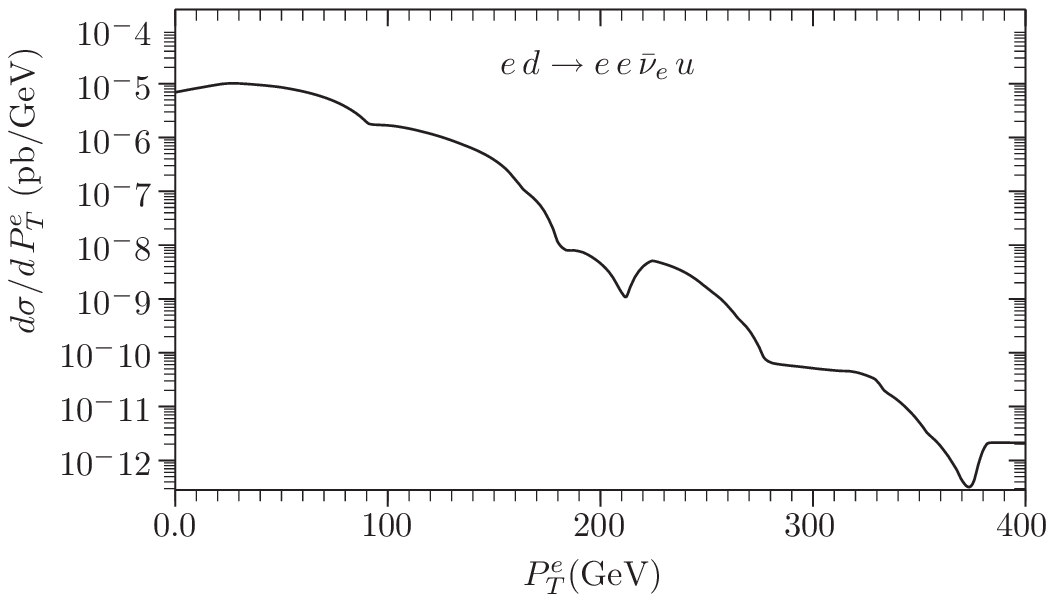}\includegraphics[width=7cm]{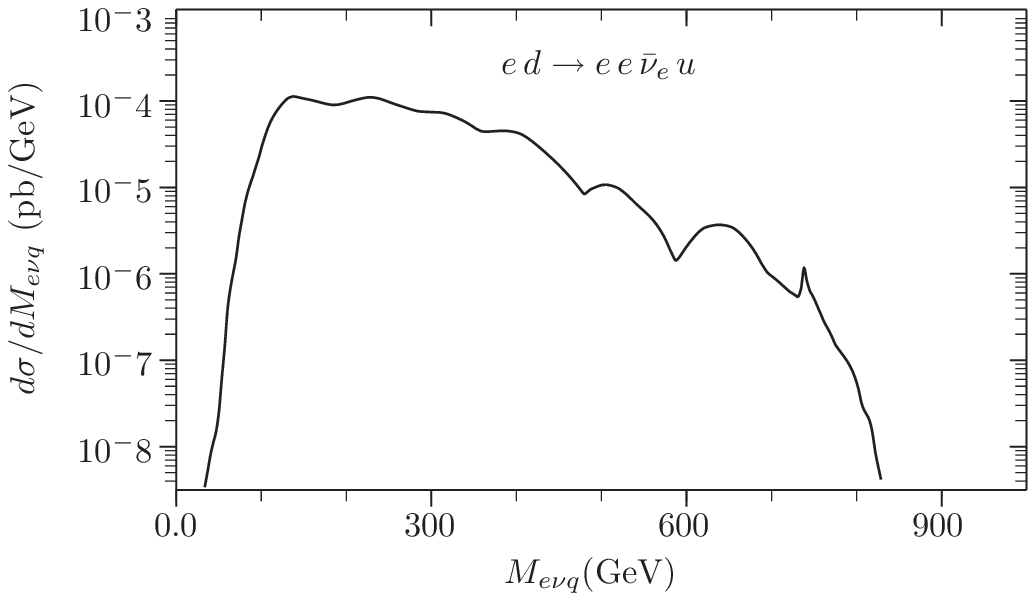}
    \includegraphics[width=7cm]{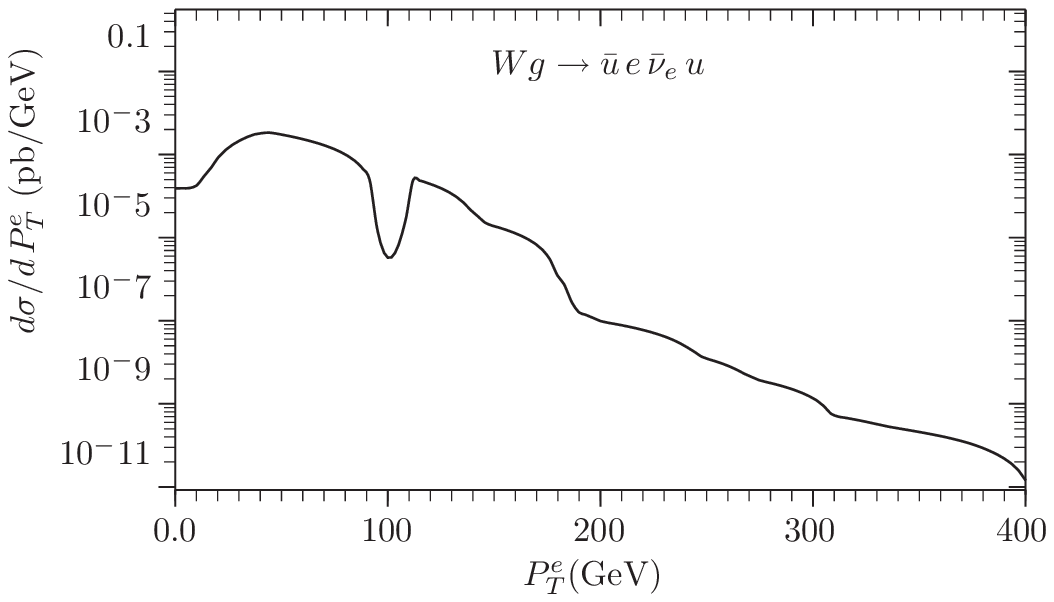}\includegraphics[width=7cm]{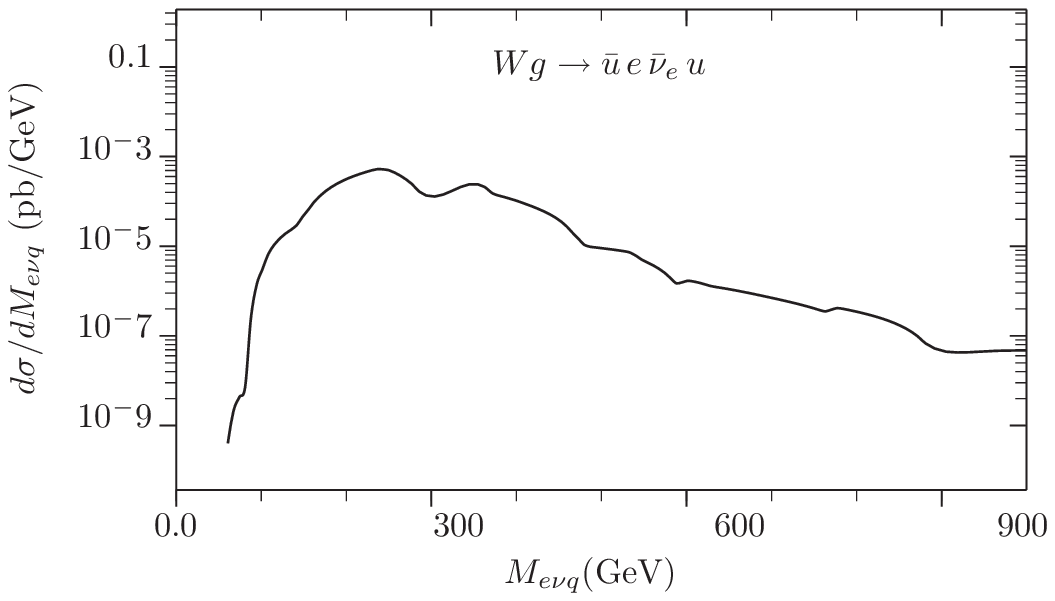}
    \includegraphics[width=7cm]{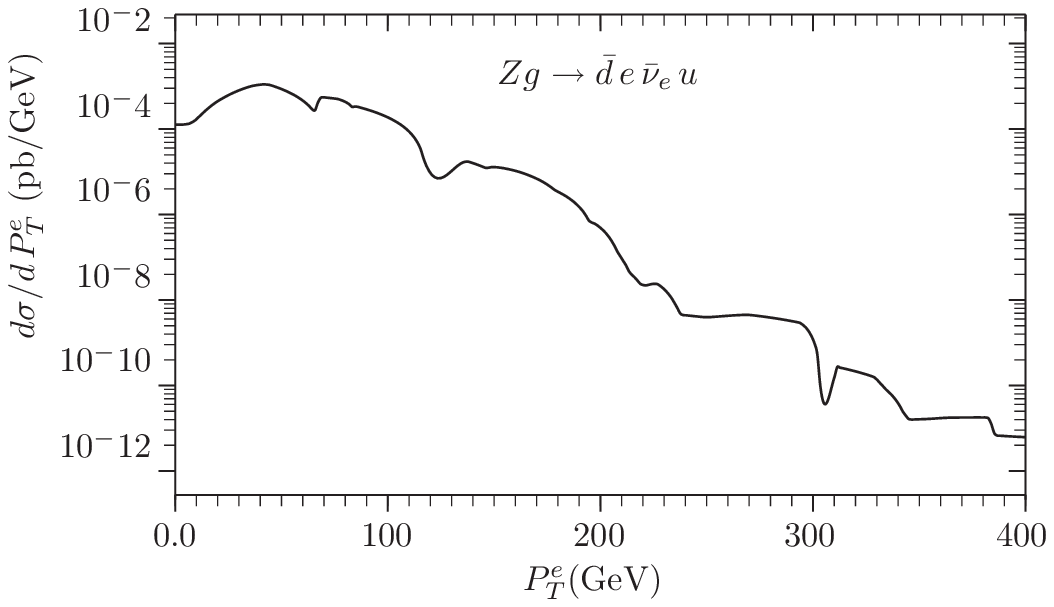}\includegraphics[width=7cm]{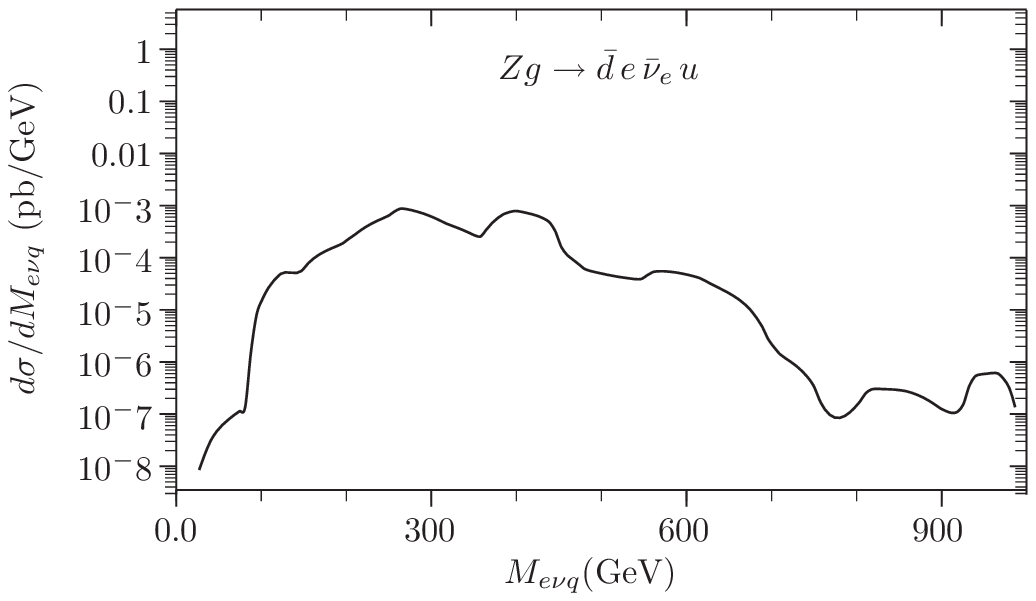}
    \includegraphics[width=7cm]{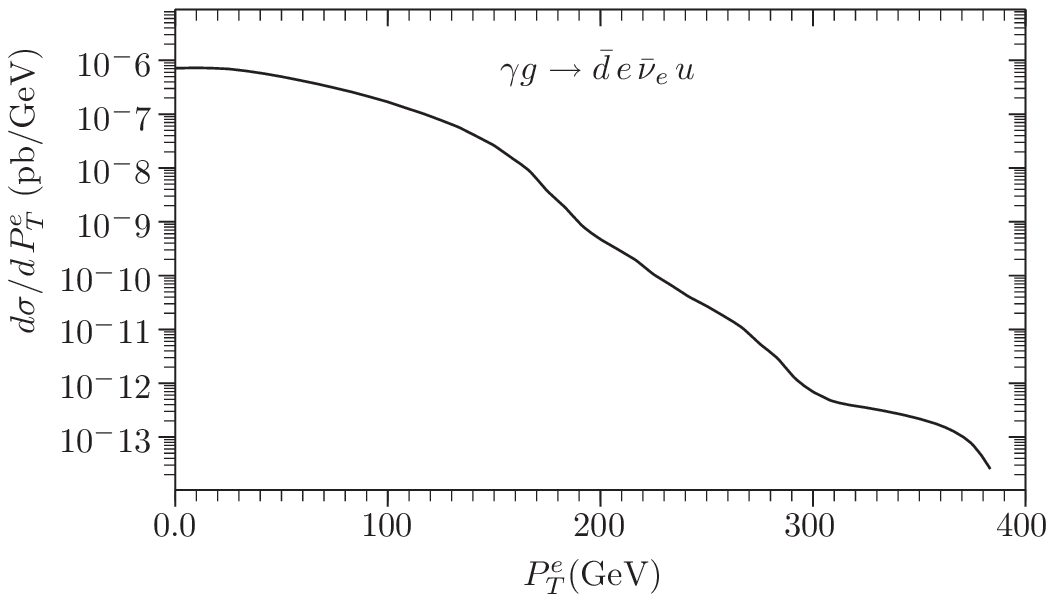}\includegraphics[width=7cm]{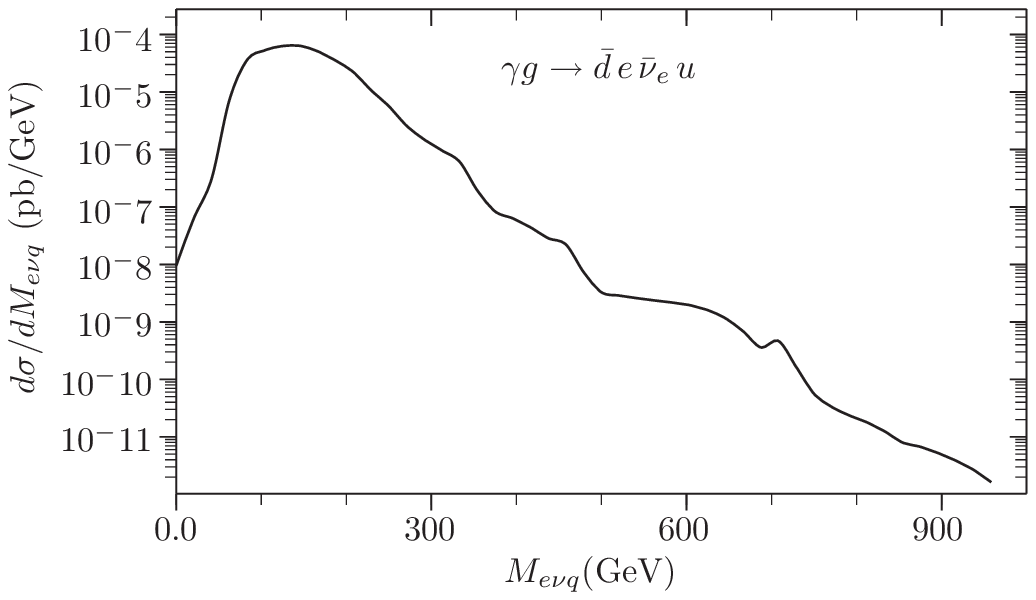}
\caption{Transverse momentum $P_T^e$ and invariant mass
distributions of the backgrounds at LHeC($\sqrt{S}$=1.4
TeV).}\label{fig8}
\end{figure}

In Figure ~\ref{fig7}, we displayed the transverse momenta, $P_T$
of secondary electrons by assigning a mass value of 350 GeV to $m$
giving peaks around $P_T^e$=30-40 GeV and the invariant mass
distributions of $e-\bar\nu_e-q$ system resulting from the decay
of $D$ quarks for the five subreactions separately. These
invariant mass spectra have Jacobian peaks of around 125-200 GeV.
Similarly, in Figure~\ref{fig8} the transverse momenta and
invariant mass distributions are displayed for LHeC energy which
yields peaks around 20-40 GeV for $P_T^e$ values and 125-350 GeV
for invariant mass values depending on different subprocess.
\section{Conclusion}

We have investigated charge -1/3 weak isosinglet vectorlike quarks
predicted by $E_6$ theories. We found that a very clear signature
for the semileptonic $e\nu_eu$ final state at two high energy
electron-proton colliders options THERA and LHeC is possible. As
presented in Tables~\ref{tab4} and~\ref{tab5} the background cross
sections varies in the range of the order of $10^{-3}-10^{-7}$
$pb$ at THERA for 500 GeV quarks and $10^{-5}-10^{-9}$ $pb$ at
LHeC for 1000 GeV quarks. As seen from Tables~\ref{tab6}
and~\ref{tab7}, exotic quark masses can be as high as 450 GeV and
1.2 TeV at the center of mass energies of $\sqrt s=1$ and 1.4 TeV,
respectively, taking $S/\sqrt{S+B}>5$, $S$ is the number of
signals and $B$ is that of background events, as an observation
limit of signal. In obtaining these numerical values, we have
taken an appropriate mixing angle value of $\sin^2\theta$=0.05
which is at the order of the angle in the CKM matrix.

\begin{acknowledgements}
This work is partially supported by Abant Izzet Baysal University
Research Fund.
\end{acknowledgements}

\newpage
\begin{table}
\centering
 \caption{The total cross sections in $pb$ for the signal processes for the exotic quarks at THERA.}\label{tab1}
\begin{tabular}{l c c c cc }
\hline\hline
$m$(GeV)&100&200&300&400&500
\\\hline
$\sigma_S$($eu\rightarrow\nu_eD$) &4.01 &2.84&1.79&0.99&0.46\\
$\sigma_S$($ed\rightarrow eD$) &0.27 &0.18&0.11&0.053&0.022\\
$\sigma_S$($Wg\rightarrow\bar uD$) &3.16 &1.25&0.39&0.10&0.019 \\
$\sigma_S$($Zg\rightarrow\bar dD$) &23.95 &9.30&2.91&0.73&0.14 \\
$\sigma_S$($\gamma g\rightarrow\bar dD$) &59.62 &9.78&1.73&0.29&0.041\\
\hline\hline
\end{tabular}
\end{table}
\begin{table}
\centering
 \caption{The total cross sections in $pb$ for the signal processes for the exotic quarks at
LHeC.}\label{tab2}
\begin{tabular}{l c c c c c c}
\hline\hline

$m$(GeV)&200&400&600&800&1000
\\\hline
$\sigma_S$($eu\rightarrow\nu_eD$) &3.61 &1.97&0.83&0.24&0.036\\
$\sigma_S$($ed\rightarrow eD$) &0.25 &0.12&0.044&0.011&0.0013\\
$\sigma_S$($Wg\rightarrow\bar uD$) &2.09 &0.45&0.061&0.0048&1.79x10$^{-4}$ \\
$\sigma_S$($Zg\rightarrow\bar dD$) &16.14 &3.49&0.47&0.037&1.42x10$^{-3}$ \\
$\sigma_S$($\gamma g\rightarrow\bar dD$) &15.16 &1.23&0.094&0.0049&1.42x10$^{-4}$\\
\hline\hline
\end{tabular}
\end{table}
\begin{table}
\centering
 \caption{The branching ratios for $D$ decays.}\label{tab3}
\begin{tabular}{c c c}
\\\hline\hline
 $m$(GeV)&BR($D\rightarrow q W$)&BR($D\rightarrow q Z$) \\\hline
150 & 70.2 \%& 29.2\%\\
250 & 68.1\%&31.9\% \\
350&  67.8\%& 32.1\%\\
450 &67.7\%&32.3\%\\
  \hline\hline
\end{tabular}
\end{table}

\begin{table}
\centering \caption{ The total cross sections for the background
processes for the exotic quarks at THERA.}\label{tab4}
\begin{tabular}{lccc c c }
\hline\hline
$m$(GeV)&100&200&300&400&500\\\hline
$\sigma_B$($eu\rightarrow\nu_e\,e\,\bar\nu_e\,u$)  &1.122x10$^{-3}$&1.451x10$^{-3}$&4.238x10$^{-4}$ &1.856x10$^{-4}$&8.976x10$^{-5}$ \\
$\sigma_B$($ed\rightarrow e\,e\,\bar\nu_e\,u$)     &8.349x10$^{-4}$&1.095x10$^{-3}$&3.332x10$^{-4}$ &8.996x10$^{-4}$&1.195x10$^{-3}$ \\
$\sigma_B$($Wg\rightarrow\bar u\,e\,\bar\nu_e\,u$) &2.011x10$^{-3}$&2.556x10$^{-2}$ &7.473x10$^{-2}$ &2.578x10$^{-2}$&1.425x10$^{-4}$ \\
$\sigma_B$($Zg\rightarrow\bar d\,e\,\bar\nu_e\,u$) & 1.502x10$^{-3}$&7.417x10$^{-3}$& 8.200x10$^{-3}$&3.843x10$^{-5}$&5.511x10$^{-4}$ \\
$\sigma_B$($\gamma g\rightarrow \bar d\, e\,\bar\nu_e\,u$) & 5.071x10$^{-3}$&1.195x10$^{-3}$&3.196x10$^{-4}$ &2.860x10$^{-5}$&7.210x10$^{-7}$ \\
  \hline\hline
\end{tabular}
\end{table}
\begin{table}
\centering
 \caption{The total cross sections in pb for
background processes for the exotic quarks at LHeC.}\label{tab5}
\begin{tabular}{lcccccc}
\hline\hline
$m$(GeV)&200&400&600&800&1000\\\hline
$\sigma_B$($eu\rightarrow\nu_e\,e\,\bar\nu_e\,u$) & 1.852x10$^{-3}$&5.002x10$^{-4}$ &2.157x10$^{-4}$&3.405x10$^{-5}$&7.724x10$^{-6}$ \\
$\sigma_B$($ed\rightarrow e\,e\,\bar\nu_e\,u$) & 8.398x10$^{-4}$&5.814x10$^{-5}$ &4.309x10$^{-5}$&4.619x10$^{-6}$&1.128x10$^{-6}$ \\
$\sigma_B$($Wg\rightarrow\bar u\,e\,\bar\nu_e\,u$) & 5.976x10$^{-3}$&5.260x10$^{-3}$&3.330x10$^{-4}$&2.352x10$^{-4}$&2.786x10$^{-5}$ \\
$\sigma_B$($Zg\rightarrow\bar d\,e\,\bar\nu_e\,u$) & 7.938x10$^{-3}$& 1.399x10$^{-2}$&1.048x10$^{-3}$&9.683x10$^{-6}$&3.436x10$^{-5}$ \\
$\sigma_B$($\gamma g\rightarrow \bar d\, e\,\bar\nu_e\,u$) &1.369x10$^{-3}$&1.090x10$^{-5}$ &1.239x10$^{-5}$&2.512x10$^{-8}$&1.114x10$^{-9}$ \\
  \hline\hline
\end{tabular}
\end{table}
\begin{table}
\centering \caption{ The significance ($S/\sqrt{S+B}$) for the
exotic quarks at THERA ($L=40$ $pb^{-1}$).}\label{tab6}
\begin{tabular}
{l c c c c c } \hline\hline
$m$(GeV)&100&200&300&400&500
\\\hline
$eu\rightarrow\nu_eD$ &12.66 &10.65&8.46 &6.29&4.30  \\
$ed\rightarrow eD$ &3.28 &2.68&2.05 &1.46&0.94 \\
$Wg\rightarrow\bar uD$ & 11.24&7.00&3.64&1.77&0.87 \\
$Zg\rightarrow\bar d D$ & 30.94&19.28&10.78&5.40&2.39 \\
$\gamma g\rightarrow \bar d D$&48.83&19.78&8.31&3.40&1.28 \\
  \hline\hline
\end{tabular}
\end{table}
\begin{table}
\centering
 \caption{The significance ($S/\sqrt{S+B}$) for the exotic quarks at
LHeC ($L=10^4$ $pb^{-1}$).}\label{tab7}
\begin{tabular}
{l c c c c c } \hline\hline
$m$(GeV)&200&400&600&800&1000
\\\hline
$eu\rightarrow\nu_eD$ &189.95 &140.33&91.04 &48.91&18.87  \\
$ed\rightarrow eD$ &49.58 &34.85&21.67 &10.35&3.54 \\
$Wg\rightarrow\bar uD$ & 144.62&66.93&24.61&6.75&1.25 \\
$Zg\rightarrow\bar d D$ & 401.64&186.42&68.62&19.32&3.72 \\
$\gamma g\rightarrow \bar d D$&389.33&111.03&30.65&7.00&1.19 \\
  \hline\hline
\end{tabular}
\end{table}


\end{document}